# From Majorana theory of atomic autoionization to Feshbach resonances in high temperature superconductors

**Alessandra Vittorini-Orgeas and Antonio Bianconi**

*Dipartimento di Fisica, Sapienza Università di Roma*
*P.le Aldo Moro 2, 00185 Roma, Italy*

The Ettore Majorana paper "Theory of incomplete P' triplets", published in 1931, focuses on the role of selection rules for the non-radiative decay of two electron excitations in atomic spectra, involving the configuration interaction between discrete and continuum channels. This work is a key step for understanding the 1935 work of Ugo Fano on the asymmetric lineshape of two electron excitations and the 1958 Herman Feshbach paper on the shape resonances in nuclear scattering arising from configuration interaction between many different scattering channels. The Feshbach resonances are today of high scientific interest in many different fields and in particular for ultracold gases and high $T_c$ superconductivity.

## 1. Introduction

Ettore Majorana was 25 years old in 1931, five years after quantum mechanics was proposed by Heisenberg and Dirac and the wave mechanics by Schrödinger. In this year he was attracted by the problem of the role of the exchange interaction in the problems of the molecular bond [1] and the decay of two electron excitations in atomic spectra [2]. The interest of the scientific community working in the development of quantum mechanics was diverging in two directions, the first, toward high energy interactions in nuclear physics, the second, toward understanding the chemical bond [3,4], and the processes of molecular association and dissociation [5-10]. It was pointed out that the dissociation involves the decay of the bound particles in a discrete state into a continuum spectrum of states made of dissociated particles. The molecular dissociation processes where considered to be similar to the Auger effect [11-18], discovered in the twenties by Lise Meitner [11,12] and Pierre Auger [13,14]. The Auger effect concerns the non-radiative decay channel of the excited states of an atom A* with the emission of a free electron leaving a positive ion

$$A^* \rightarrow A^+ + e^-_{auger}$$

This decay process competes with the standard radiative recombination, where a neutral excited atom A* atom decays in its ground state A with the emission of a photon

$$A^* \rightarrow A + h\nu.$$

In 1921 Albert Einstein received the Nobel prize for the theoretical interpretation of the photoemission process

$$h\nu + A \rightarrow A^+ + e^-_{photoelectron}$$

that involves the excitation of an electron in the external atomic shell in the vacuum so that the ion $A^+$ is left in its ground state.

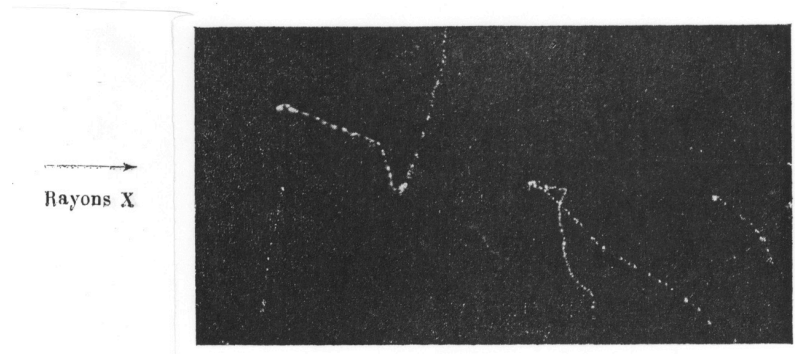

*Fig. 1 The original photographic plate where Auger has recorded two electrons emitted by a krypton atom that have absorbed an x-ray photon of 60 KeV energy in a Wilson chamber. The Auger experiment shows that by changing the energy of the x-ray photon the energy of the Auger electron does not changes while the energy of the photoelectron increases linearly with the photon energy (1926) [14].*

The Auger effect was found looking at the photoemission process in the x-ray range where two electrons are emitted, as it can be seen in Fig. 1. In this energy range the photons have enough energy to excite a photoelectron from a core level (for example the 1s level, called K shell) of an atom A into the vacuum. The ion, with N-1 electrons, is formed in an excited state $A^{+*}$ (for example with a core hole in the 1s level or K shell). This excited state can decay via a radiative x-ray fluorescence process, known as x-ray fluorescence channel (for example the $K_\alpha$ x-ray emission line following the electronic transition from 2p to 1s level)

$$h\nu + A \rightarrow e^-_{photoelectron} + A^{+*} \rightarrow e^-_{photoelectron} + A^+ + h\nu_{fluorescence}.$$

Pierre Auger discovered that this process is in competition with the non-radiative decay of $A^{+*}$

$$h\nu + A \rightarrow e^-_{photoelectron} + A^{+*} \rightarrow e^-_{photoelectron} + A^{2+} + e^-_{auger}$$

where the core hole in the inner level is filled by an electron in a more external level with the emission of an Auger electron. Auger described this process as the dissociation of the excited $A^{+*}$ into a $A^{2+}$ ion (with N-2 electrons) and an Auger electron in the continuum following the energy scheme shown in Fig. 2.

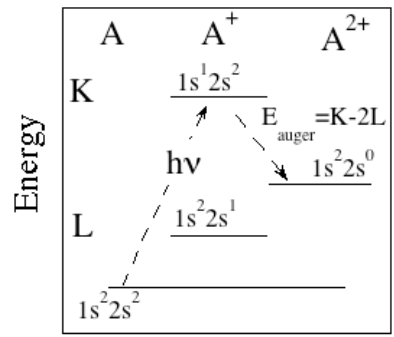

Fig. 2. The energy scheme for the case of the excitation of a core hole in the 1s level (K shell) followed by the emission of a photoelectron and the emission of one Auger electron of energy K-2L. $h\nu + A(1s^2 2s^2) \rightarrow e^-_{photoelectron} + A^{+*}(1s^1 2s^2) \rightarrow e^-_{photoelectron} + A^{2+}(1s^2 2s^0) + e^-_{auger}$.

In 1927 Wentzel [19] proposed a non trivial extension of the Heisenberg theory of the quantum resonance in helium published in 1926 [20] to explain the Auger effect. He considered the simple case of the non-radiative decay of a discrete excited state of two electrons into final states in the continuum. He proposed that the Auger decay was due to the configuration interaction, driven by the Coulomb repulsion, between the initial discrete excited state of two electrons and the continuum final state where one of two electrons is emitted as a free electron in the vacuum. The wavefunction of the initial state of the two excited electrons of quantum numbers $n, m, l$ and $N, M, L$ and coordinates $r, \theta, \varphi$ and $R, \Theta, \Phi$ respectively was described as the linear combination of the products of the hydrogenoic wavefunctions $f_{nlm}$ e $f_{NLM}$:

$$u_{nlmNLM} = \frac{1}{\sqrt{2}} \{ f_{nlm}(r\theta\varphi) \cdot f_{NLM}(R\Theta\Phi) \pm f_{nlm}(R\Theta\Phi) \cdot f_{NLM}(r\theta\varphi) \}.$$

The effect of the perturbation is to mix with different weights the wavefunctions of the system and it is described by

$$\sum_{n'l'm'L'M'} \int_0^\infty dE' \frac{V_{nlmNLM}^{n'l'm'E'L'M'} u_{n'm'l'E'L'M'}}{(E_n - E_{n'}) - (E' - E_N)}$$

where the sum has been replace by an integral considering a continuum spectrum

The term $V_{nlmNLM}^{n'l'm'E'L'M'}$ represents the matrix associated with the transition from the initial discrete states $nlm$ and $NLM$ to the final $n'l'm'$ and $N'L'M'$ states in the continuum.

$$V_{nlmNLM}^{n'l'm'E'L'M'} = \int d\varphi \int d\theta \, sen\theta \int dr \, r^2 \int d\Phi \int d\Theta \, sen\Theta \int dR \, R^2$$
$$\cdot \frac{e^2}{|R - r|} u_{nlmNLM} \cdot u_{n'l'm'E'L'M'}$$

where the $e^2/|R-r|$ is the Coulomb interaction. The approximate wavefunction of the final states in the continuum has been described as:

$$f_{E'L'M'}(r\theta\varphi) = \sqrt{\frac{2}{\pi} \frac{dk'}{dE'}} \cdot \frac{\cos(k'r - \alpha)}{r} \cdot P_{L'M'}(\theta\varphi), \text{ where } k' = (2\pi/h)\sqrt{2mE'}$$

with a time dependent factor $e^{\hbar i(E - mc^2)t} \cdot e^{-\gamma t}$, where $e^{-\gamma t}$ takes into account the decrease of the electronic charge due to the electron emission in the vacuum.

Although the possibility of the extension of the quantum resonance between two bound states to that between a discrete and a continuum of a many body system could be questioned, Wentzel showed that his theory was able to predict an important experimental fact: the ratio between the radiative and non-radiative recombination as a function of atomic number Z. The calculated probability of the Auger non-radiative recombination channel is larger (lower) than that for the radiative recombination for low (high) Z atoms in agreement with experimental results [15-18].

The absorbed photon energy in the non-radiative decay of the photo-excited states A* is fully converted into an electronic energy, with no emission of photons, therefore it is of high interest for the efficient transformation of photonic energy into electronic energy, but this quantum process remains difficult to be predicted theoretically since it depends on the details of the quantum overlap of the many bound and unbound electronic wave functions involved in the process.

Today the non-radiative decay is of high scientific interest in the photosynthesis that is a key process to understand living matter. In fact it appears that the biological evolution has selected particular quantum tricks for the control of the non-radiative decay of the photo-excited state A* in the antenna of the photosystyem II to get the most efficient conversion of the absorbed electromagnetic radiation into an electronic energy: the voltage gradient across the thylakoid membrane called the proton-motive force.

## 2. Two electron excitations

Ettore Majorana [2] addressed his interest to the selection rules for the non-radiative recombination process of two electron excited states following the Wentzel theory. The probability of the decay of the two electron quasi-bound excited state into the continuum is controlled by the Coulomb interaction in the Wentzel theory. Majorana pointed out in his work that it was necessary to consider also the configuration interaction due to spin orbit interaction between the atomic two electron excited states in order to control the probability of the non-radiative decay.

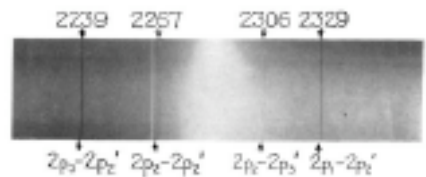

*Fig. 3. The two electrons excitations in the optical absorption spectra of Cd measured by Foote et al. (1925) [21].*

He recognized that the ideal simplest experiment for testing the quantum theory for the decay of the two electron excited states, considered in the Wentzel theory, was the decay of the excited state of a neutral atom with two external electrons measured in the optical spectra- where a photon excites the two external electrons.

These two electron excitations were measured as week absorption lines in the atomic spectra beyond the ionization potential (I.P.) for single electron excitations [21-23]. The optical absorption spectrum of Cd, measured by Foote et al. in 1925 [21] is shown in Fig. 3. In this experiment the absorption spectra of atomic species with two external electrons $ns^2$ have been recorded. The final states for the two electron excitations $ns^2np^0 + h\nu \rightarrow ns^0np^2$, where both electrons are excited into the $p$ level, have been identified. The energy scheme of the two electron excitations observed in Cd, Zn and Hg is shown in Fig. 4. The quasi-bound discrete states of the two excited electrons, $np^2$, are degenerate with the continuum and therefore they are expected to decay with the Wentzel non radiative recombination if the process is not forbidden by selection rules.

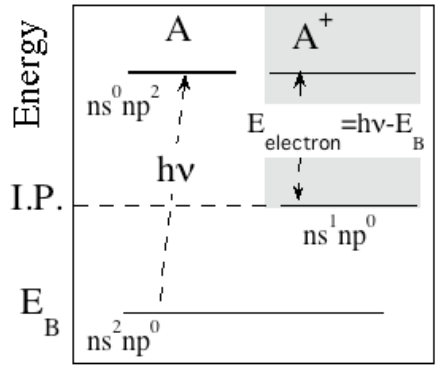

*Fig. 4. The energy scheme for two electron excitations in Cd, Zn and Hg and the degenerate continuum states involved in the non radiative decay process.*

Majorana has shown in his work [2] that the expected final states are

$$\begin{cases} J = 3: \ (^3D_3) \\ J = 2: \ ^1D_2, \ ^3P_2, \ (^3D_2) \\ J = 1: \ ^3P_1, \ (^3D_1), (^1P_1), (^3S_1) \\ J = 0: \ ^3P_0, \ ^1S_0. \end{cases}$$

The final states $(^3D_3), (^3D_2), (^3D_1), (^1P_1), (^3S_1)$ are forbidden by the Pauli principle, and the $^3P_0$ $^3P_1$ $^3P_2$ final triplet states are expected to be observed, on the contrary the $^3P_2$ was not observed in all spectra of Cd, Zn, and Hg spectra. This unexpected experimental result was known as the problem of the incomplete P' triplet. Majorana has showed in his paper that it is

A. Vittorini-Orgeas and A. Bianconi arXiv:0812.1551 (2008)

necessary to consider the configuration interaction of the excited states induced by spin orbit interaction that gives:

$$\begin{cases} H\,^3P_2 = \dfrac{f(r_1)+f(r_2)}{4}{}^3P_2 + \dfrac{f(r_1)+f(r_2)}{2\sqrt{2}}{}^1D_2 + \sqrt{3}\,\dfrac{f(r_1)-f(r_2)}{4}({}^3D_2) \\ H\,^3P_1 = -\dfrac{f(r_1)+f(r_2)}{4}{}^3P_1 + \sqrt{5}\,\dfrac{f(r_1)-f(r_2)}{4\sqrt{3}}({}^3D_1) + \\ \qquad + \dfrac{f(r_1)-f(r_2)}{2\sqrt{2}}({}^3P_1) - \dfrac{f(r_1)-f(r_2)}{\sqrt{3}}({}^3S_1) \\ H\,^3P_0 = -\dfrac{f(r_1)+f(r_2)}{2}{}^3P_0 + \dfrac{f(r_1)+f(r_2)}{\sqrt{2}}{}^1S_0 \end{cases}$$

Majorana results show that the stable terms are due to even $4p^2$ configurations which cannot decay into the odd continuum $4s^1,\varepsilon p$ configuration. Though partial breakdown of LS coupling in the $p^2$ configuration the unstable term ($^3P_2$) acquires a component of ($^1D_2$) and thus become subject to allows the non radiative decay through interaction with the $4s,\varepsilon d$ continuum. Therefore the $^3P_2$ term is the only level in configuration the J=1 subject to rapid non radiative decay because of its interaction with the final state $^1D_2$. Therefore Majorana has shown that the spin orbit interaction need to considered for the control of the non-radiative recombination and in the particular experimental case he considered it becomes relevant only for the missing term in the spectrum.

Clearly for testing the predictions of the quantum mechanics the most simple experiment on two electron excitations is the absorption spectrum of atomic helium beyond the ionization potential. Unfortunately the two electron excitations in helium are in the ultraviolet energy range and they were not available in 1931 for Majorana.

The ultraviolet absorption spectrum of helium will be measured few years later in 1935 by H. Beutler [24] who identified the series of two electron excitations $1s^2 + h\nu \rightarrow 1s^0 2s^1 np^1$ shown in Fig. 5.

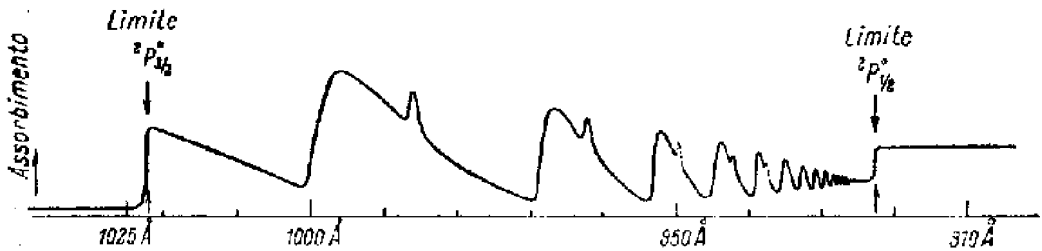

Fig. 5. The ultraviolet absorption spectrum of helium beyond the ionization limit measured by H. Butler (1935)[24].

The paper of H. Butler was known at the Institute of Physics of Rome University in via Panisperna in January 1935. The Fermi's group was no more working on basic problems of quantum mechanics, they were fully focused on the experiments in the artificial radioactivity induced by the nuclear capture of slow neutrons discovered few month before. Emilio Segrè were helping Enrico Fermi in his experimental efforts to investigate the nuclear capture in many different atomic species, but Ettore Majorana was working at home in Rome and was not coming to the Institute in via Panisperna. Emilio Segrè kept the contacts with Ettore and visited Ettore at home in viale Regina Margherita in Rome several times, as Amaldi has written in his memories [25].

Ugo Fano, coming form Turin with a diploma in mathematics, joined the Fermi's group at the end of 1934. Ugo Fano has written his memories of that period [26] "In January, 1935, Emilio Segrè gave me some spectroscopy papers by H. A. Beutler as a fruitful subject of study. The Beutler spectra showed unusual intensity profiles which struck me as reflecting interferences between alternative mechanisms of excitation. Fermi then taught me sequentially within a few days how to formulate my interpretation theoretically; a paper was sent to the *Nuovo Cimento* quickly and I *dropped the matter*."

It is possible therefore that E. Segrè was the link between Ettore Majorana and Ugo Fano in fact it is clear that the 1935 Fano paper [26] is the evolution of the 1931 Ettore Majorana paper [2]. The Panisperna group, Enrico Fermi, Emilio Segrè, Edoardo Amaldi, like Ugo Fano, "*dropped the matter*" and considered this work as a simple exercise of quantum mechanics in atomic physics of no interest in nuclear physics and other fields.

The paper by Ugo Fano identifies the anisotropic lineshape of the profile of the two electron absorption lines as due to the configuration interaction between the quasi bound excited state and the continuum. The anisotropy coefficient can be extracted from the experimental data and it provides a direct measure of the complex integral of configuration interaction between two different scattering channels that is very difficult to calculate for the specific cases.

## 3. Synchrotron radiation spectroscopy

In the early 60' the revival of the interest on two electron excitations in atomic spectra was due to the discovery of the synchrotron radiation emitted by electron synchrotrons that because of its intense continuum spectrum extending from the infrared to x-ray range allowed for the first time to extend the experimental spectroscopy investigation to far ultraviolet and soft x-ray range. Synchrotron radiation spectroscopy experiments beyond the near ultraviolet started at the National Bureau of Standards (NBS) in USA where in these years Ugo Fano was working. Ugo Fano published in 1961 an extended version of his 1935 paper in Phys. Rev. [28] that had an impressive success and today it classified between the most important publications in the physics of the second half of the XX century (this paper is between the first three most relevant works published in Phys. Rev. and Phys. Rev. Lett.). The success of this theory has been related with the emergence of synchrotron radiation spectroscopy that greatly expanded the range of

A. Vittorini-Orgeas and A. Bianconi arXiv:0812.1551 (2008)

spectroscopy from the optical or near ultraviolet to far ultraviolet in the early 1960s, to the soft x-rays in the 1970s and in the hard x-rays in the 1980s. In the early 1960s the development of a synchrotron radiation source at the betatron of the National Bureau of Standards (NBS) was an offshoot of a program of investment in betatron electron accelerators which started in the late 1940s, originally aimed to provide high-energy electron beams for the production of x rays. These accelerators also yielded, at first as an unappreciated byproduct, a broad band of synchrotron radiation spanning the electromagnetic spectrum from the radio to the x-ray domains. The first relevant experiment with synchrotron radiation was the measure in 1963 of the two electron excitations in the far ultraviolet [29-31] of noble gases shown in Fig. 6. The experimental absorption profile of the two electron excitations shown in Fig. 7 are in fully agreement with the theory of configuration interaction between continuum and discrete states The two electrons excitations have been investigated later at other synchrotron radiation facilities in the soft x-ray range in the solids [32,33] and molecules [34] and later in the x-ray range in solids [35] using novel synchrotron radiation facilities using the storage ring. The x ray spectroscopy with synchrotron radiation in these last 40 years allowed a careful test of the theory of configuration interaction between a discrete and a continuum [36].

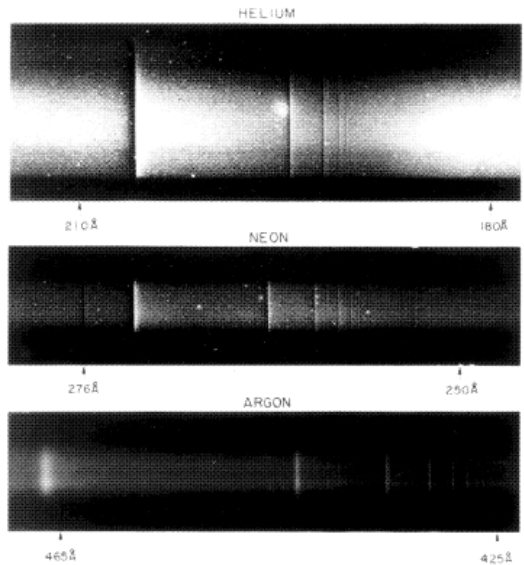

*Fig. 6. Absorption spectra of helium, neon, and argon atoms in the extreme ultraviolet spectral region, from [29-31]. These images have been recorded on photographic plates exposed to synchrotron radiation at the National Bureau of Standards in USA. The synchrotron radiation was passed through a gas cell and then dispersed by a diffraction grating to show the dependence of absorption upon wavelength (which is indicated in Ångstrom units: 1 Å = $10^{-10}$ m). Increased blackness indicates increased absorption by the gas. Note that these spectral lines clearly exhibit the Fano asymmetric profiles.*

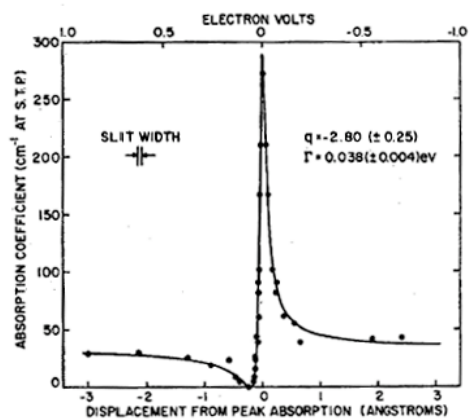

*Fig. 7. The absorption coefficient vs. wavelength for the strongest absorption feature of helium (around 206 Å). The points are experimental data; the solid line is a fit to the Fano profile formula. This feature is associated with excitation of the $2s2p$ $^1P_0$ state of the helium atom (1963) from ref. [31].*

## 4. Shape resonances in nuclear physics

In the fifties the interest of nuclear physics was attracted by the resonant scattering of a slow neutron on a nucleus known as the "shape resonances" [37]. In 1954, Herman Feshbach, Charles Porter, and Victor Weisskopf [38] developed the cloudy crystal ball model, which revolutionized the treatment of nuclear reactions, initially providing a detailed description of the scattering of neutrons from nuclei. The model's characterization of the nucleus as a complex "optical" potential combines the independent nucleon aspect embodied in the shell model with the excitation of dense compound nuclear levels postulated by Niels Bohr to explain many aspects of nuclear reactions. Herman Feshbach developed the theory of the multichannel resonance [39] in the frame of the general nuclear reaction theory for multistep reactions (known today as Feshbach resonance). In 1961 Fano [28] realized that the Feshbach theory was an extension of his paper in Nuovo Cimento in 1935. In fact the Feshbach theory deals with neutron capture that is an association processes inverse to those considered in the Fano's theory of autoionization that is a dissociation process. The neutron capture is a scattering process, in which the system is formed by combining an incident particle n (the neutron) with the "rest" (the nucleus). Moreover in nuclear multi-channel reactions the nucleus (the system of N+1 particles) then breaks up releasing alternatively either the same particle or another particle and …, therefore the multichannel Feshbach resonance was an extension of the single channel shape resonance discussed in the Fano's paper of 1935. The theory of shape resonances and multichannel Feshbach resonances was developed and applied to many processes of association and dissociation [40].

## 5. Exchange interaction in superfluids at high temperature

Feshbach resonances for molecular association and dissociation have been proposed for the manipulation of the interatomic interaction in ultracold atomic gases. In fact the interparticle interaction shows resonances tuning the chemical potential of the atomic gas around the energy of a discrete level of a biatomic molecule controlled by an external magnetic field [41-43]. This quantum phenomenon has been used by Ketterle group to achieve the Bose-Einstein condensation (BEC) in the dilute bosonic gases of alkali atoms [44] and by Jin's group to reach the crossover between BCS and Bose condensation in fermionic gases [45] where Feshbach resonances control the exchange pairing between fermions [46-49].

In the field of high $T_c$ superconductivity the process for increasing $T_c$ by "shape resonance" was first proposed by Blatt and Thompson [50] in 1963 for a superconducting thin film and extended later to a superlattice of quantum stripes [51-54] or quantum wells [55].

In conclusion we have shown that the 1931 paper of Majorana [2] plays a key role for understanding the 1935 theory of Fano for the autoionization lines of helium involving the configuration interaction between discrete and continuum ionization channels that is at the basis of Feshbach resonances in many body systems.

.

**References**

[1] E. Majorana, "Sulla formazione dello ione molecolare di He", *Nuovo Cimento*, **8** (1931) 22.

[2] E. Majorana, "Teoria dei tripletti P' incompleti", *Nuovo Cimento*, **8** (1931) 107.

[3] W. Heitler, F. London, "Wechselwirkung neutraler Atome und homopolare Bindung nach der Quantenmechanik", *Zeitschrift für Physik*, **44** (1927) 455.

[4] R. Pucci1 and G. G. N. Angilella, "Majorana: From Atomic and Molecular, to NuclearPhysics" Foundations of Physics (2006) DOI: 10.1007/s10701-006-9067-7

[5] R. de L. Kronig, *Zeitschrift für Physik*, **62** (1930) 300.

[6] M. Polanyi and E. Wigner *Zeitschrift für Physik*, **33** (1925) 429.

[7] O. K. Rice *Phys. Rev.,* **33** (1929) 748.

[8] O. K. Rice *Phys. Rev.,* **35** (1930) 1538; ibidem **35** (1930) 1551.

[9] O. K. Rice *Phys. Rev.,* **37** (1931) 1187.

[10] O. K. Rice *Phys. Rev.,* **38** (1931) 1943.

[11] L. Meitner, *Zeitschrift für Physik,* **9** (1922) 131.

[12] L. Meitner, *Zeitschrift für Physik,* **17** (1923) 54.

[13] P. Auger, *Comptes rendus,* **180** (1925) 65.

[14] P. Auger, *Journal de Physique,* **6** (1925) 205.

[15] P. Auger, *Ann. de Phys.* **6** (1926) 183.

[16] M. de Broglie und J. Thibaud, *Comptes Rendus* **180** (1925) 179.

[17] H. R. Robinson, *Nature*, **118** (1926) 224

[18]  H. R. Robinson and A. M. Cassie, *Proc. Roy. Soc. (A)* **113** (1926) 282

[19] G. Wentzel, *Zeitschrift für Physik,* **43** (1927) 524.

[20] W. Heisenberg, *Zeitschrift für Physik,* **38** (1926) 411.

[21] P. D. Foote, T. Takamine, R. L. Chenault, *Phys. Rev.,* **26** (1925) 165.

[22] E. O. Lawrence, *Phys. Rev.,* **28** (1926) 947.

[23] G. Shenstone, *Phys. Rev.,* **28** (1931) 873.

[24] H. Beutler, *Zeitschrift für Physik,* **93** (1935) 177.

[25] E. Amaldi, *Giornale di Fisica* **9** (1968) 300.

[26] U. Fano *Citation Classics,* vol. **27,** July 4 (1977).

[27] U. Fano, condmat/ 0502210.

[28] U. Fano, *Phys. Rev.,* **124** (1961) 1866.

[29]  R. P. Madden and K. Codling, *Phys. Rev. Lett.,* **10** (1963) 516

[30] J. W. Cooper, U. Fano, and F. Prats, *Phys. Rev. Lett.* **10**, (1963). 518

[31] R. P. Madden and K. Codling, *Astrophys. J.* **141**, (1965) 364.

[32] A. Balzarotti, A. Bianconi, and E. Burattini, *Phys. Rev.* B **9** (1974) 5003.

[33] A. Balzarotti, A. Bianconi, E. Burattini, M. Grandolfo, R. Habel, and M. Piacentini *Physica Status Solidi (b)* **63** (1974) 77.

[34] A. Bianconi, R. Z. Bachrach, H. Petersen, and F. C. Brown *Physical Review A* **17** (1978) 1907.

[35] A. V. Soldatov, T. S. Ivanchenko, S. Della Longa, and A. Bianconi *Phys. Rev. B* **47** (1993) 16155; A. Bianconi, J. Garcia, M. Benfatto, A. Marcelli, C. R. Natoli, and M. F. Ruiz-Lopez *Phys. Rev. B* **43** (1991) 6885; I. Davoli, A. Marcelli, A. Bianconi, M. Tomellini and M. Fanfoni Phys. Rev. B33, 2979, (1986)

[36] A. Bianconi in *X Ray Absorption: Principle, Applications Techniques of EXAFS, SEXAFS and XANES* edited by R. Prinz and D. Koningsberger, J. Wiley and Sons, New York 1988.

[37] J. M. Blatt, and V. F. Weisskopf, *Theoretical Nuclear Physics*, John Wiley & Sons Inc. New York (1952), pp. 379 ff. in particular p. 401.

[38] H. Feshbach, C. E. Porter, and V. F. Weisskopf *Phys. Rev.*, **96** (1954) 448.

[39] H. Feshbach, *Ann. Phys. (N.Y.)* **5** (1958) 357.

[40] Ugo Fano *"Atomic Collisions and Spectra"* Elsevier, Amsterdam (1986).

[41] E. Tiesinga, B.J. Verhaar, and H.T.C. Stoof., *Phys. Rev. A* **47** (1993) 4114.

[42] A. J. Moerdijk, B. J. Verhaar, amd A. Axelsson, *Phys. Rev. A* **51** (1995) 4852.

[43] R. C. Forrey, N. Balakrishnan, V. Kharchenko, and A. Dalgarno *Phys. Rev. A* **58** (1998) R2645.

[44] S. Inouye, M. R. Andrews, J. Stenger, H.-J. Miesner, D. M. Stamper-Kurn and W. Ketterle *Nature* **392** (1998) 151.

[45] C. A. Regal, M. Greiner, and D. S. Jin, *Phys. Rev. Lett.*, **92** (2004) 040403.

[46] C. H. Schunck, M. W. Zwierlein, C. A. Stan, S. M. F. Raupach, and W. Ketterle, A. Simoni, E. Tiesinga, C. J. Williams, and P. S. Julienne *Phys. Rev. A,* **71** (2005) 045601.

[47] J. Werner, A. Griesmaier, S. Hensler, J. Stuhler, T. Pfau A. Simoni and E. Tiesinga *Phys. Rev. Lett.,* **94** (2005) 183201.

[48] C. H. Schunck, M. W. Zwierlein, C. A. Stan, S. M. F. Raupach, and W. Ketterle, A. Simoni, E. Tiesinga, C. J. Williams, and P. S. Julienne *Phys. Rev., A* **71** (2005) 045601.

[49] J. Werner, A. Griesmaier, S. Hensler, J. Stuhler, T. Pfau A. Simoni and E. Tiesinga *Phys. Rev. Lett.,* **94** (2005) 183201.

[50] J. M. Blatt and C. J. Thompson *Phys. Rev. Lett.,* **10** (1963) 332.

[51] A. Bianconi and M. Missori *The Coupling of a Wigner Charge Density Wave with Fermi Liquid from the Instability of a Wigner Polaron Crystal: A Possible Pairing Mechanism in High $T_c$ Superconductors* in "Phase Separation in Cuprate Superconductors", E. Sigmund and K. A. Müller, editors Proc. of the Workshop held at Cottbus, Germany Sept 4-10, 1993, Springer Verlag, Berlin-Heidelberg (1994). 272.

[52] A. Perali, A. Bianconi, A. Lanzara, and N. L. Saini *Solid State Commun.,* **100** (1996) 181.

[53] A. Bianconi, A. Valletta, A. Perali, and N. L. Saini, *Solid State Commun.,* **102** (1997) 369.

[54] A. Bianconi N. L Saini, A. Lanzara, M. Missori, T. Rossetti, H. Oyanagi, H. Yamaguchi, K. Oka, and T. Ito *Phys. Rev. Lett.,* **76** (1996) 3412.

[55] A. Bianconi, D. Di Castro, S. Agrestini, G. Campi, N. L. Saini, A. Saccone, S. De Negri and M. Giovannini *J. Phys.: Condens. Matter* **13** (2001) 7383.

[56] A. Bianconi *"Feshbach shape resonance in multiband superconductivity in heterostructures"* Journal of Superconductivity" **18** (2005) 25.